# Scaling of Wall Turbulence at Finite Reynolds Number

Dmitrii Ph. Sikovsky

[1]Institute of Thermophysics SB RAS, Novosibirsk,630090, Russia
[2] Novosibirsk State University, Novosibirsk, 630090, Russia

**Abstract**. The scaling in the near-wall region of turbulent wall-bounded flows has long been the source of polemics, especially because of the issues related to the complete or incomplete Reynolds number similarity of the flow. In this paper the new scaling relations for the mean velocity and Reynolds shear stress in viscous sublayer were proposed based on the application of matched asymptotic expansion method to the mean momentum balance. It was shown that the new parameter $\Lambda = -\langle uv \rangle / y^3 \big|_{y=0}$ is relevant for the wall turbulence in addition to the friction velocity. From the proposed new scaling of the viscous sublayer the power law for the mean velocity in the overlap region was derived. The complete or incomplete similarity of the parameter $\Lambda_+$ was shown to be the key factor, which determines whether the mean velocity obeys the power law or log law in the overlap layer.

The scaling of wall turbulence at high Reynolds numbers has been in the focus of research beginning from the seminal works of L.Prandtl, G.I.Taylor, Th. von Kármán, A.A.Izakson, C.B.Millikan (see [1] and references therein). For turbulent flows in channels, pipes or zero pressure gradient turbulent boundary layers (ZPGTBL) the classical scaling is based on two main assumptions [2]. The first assumption is that the wall turbulence has a two-layer structure, in which the outer layer occupies almost all of the flow except the thin near-wall inner layer, or viscous sublayer, directly affected by fluid viscosity. According to Townsend's principle of Reynolds number similarity [3], the direct action of viscosity on the mean flow in the outer layer is negligible. The second assumption is that the friction velocity $v_\tau$ is the only scale for both the velocity fluctuations and mean velocity difference in the viscous sublayer. From these assumption the law-of-the-wall for the mean velocity in the viscous sublayer $U = v_\tau U_+(v_\tau y / \nu)$ can be derived on dimensional grounds as well as the von Kármán defect law $U_e - U = v_\tau F(y / \delta)$ for the outer layer. Here $U_e$ is the centerline velocity for a channel (pipe) flow or the outer edge velocity for ZPGTBL, $\delta$ is the outer lengthscale (the channel half-width, pipe radius or boundary layer thickness), and $\nu$ is the fluid kinematic viscosity.

Following the well-known Izakson-Millikan arguments, both these expressions are assumed to be valid in the overlap region $\nu/v_\tau \ll y \ll \delta$ between inner and outer layers, which results in the logarithmic law for the mean velocity $U_+ = \kappa^{-1} \ln y_+ + B$ for $y_+ \gg 1$ and $U_+ - U_{e+} = \kappa^{-1} \ln \eta + B'$ for $\eta \ll 1$. Here the von Kármán $\kappa$ and the additive $B, B'$ 'constants' are supposed to be universal, the subscript (+) denotes the dimensionalization with the velocity $v_\tau$ and kinematic viscosity $\nu$ (wall units) and $\eta = y/\delta$ is the stretched coordinate for the outer layer.

However, extensive and careful experimental studies of canonical wall-bounded turbulent flows have shown the lack of universality of the coefficients $\kappa$, $B$, $B'$ (see [1,4]); for example, the experimental value of $\kappa$ varies from 0.368 to 0.421 for different flows. Such uncertainties have given rise to a recent debate on the universality of the log law; several new scalings have been developed for the turbulent wall-bounded flows justifying the power law as an alternative to the log law [5–9]. The recently revealed influence of the very large-scale vortical structures of the outer layer on the statistics of the near-wall turbulence, especially, the Reynolds number dependence of the streamwise Reynolds stress may also prejudice the uniqueness of the friction velocity as the only scale for the turbulence in a viscous sublayer [10].

The method of matched asymptotic expansions (MAE) is the most suitable method for the analysis of wall turbulence due to its two-later structure as it is mentioned in the comprehensive review [11]. However, the previous works used MAE procedure (see [11] and references herein) were based on the assumption that all turbulent fluctuations scaled with the friction velocity. In the present letter the MAE method is applied to the problem of the scaling of wall turbulence on the base of the new inner scaling proposed. This new scaling gives a better collapse of the DNS data for both mean velocity and Reynolds shear stresses profiles in the viscous sublayer at a wide range of Reynolds numbers.

The starting point is the momentum balance for the turbulent flow in a channel, pipe, or TBL, which can be written in the following form

$$-\langle uv \rangle = v_\tau^2 + \alpha y + \int_0^y D_t U dy - \nu U_y, \qquad (1)$$

where $\alpha = \rho^{-1} dP/dx$ is the kinematic pressure gradient, $\langle uv \rangle$ is Reynolds shear stresses, $\nu$ is fluid kinematic viscosity, $D_t U = UU_x + VU_y$ is the convective terms. Based on (1) the outer asymptotic expansions of $U$ and $\langle uv \rangle$ in powers of $\nu$ can be written as

$$-\langle uv \rangle = v_\tau^2 + \alpha y + \int_0^y D_t U dy + O(\nu),$$
$$U = U_e + u_o U_1(\eta; x) + O(\nu), \qquad (2)$$

where $u_o$ is outer scale of velocity defect, $U_1$ is the order one function. The leading-order approximation for Reynolds shear stress (2) cannot satisfy the zero boundary condition at the wall owing to a singular perturbation type of the problem. Applying MAE method we have to introduce the near-wall inner layer, or viscous sublayer, with the stretched coordinate $y_\nu = y/l_\nu$, where $l_\nu$ should be chosen so that the viscous stresses would be of the same order as the Reynolds stresses. To find the expression for $l_\nu$ it is convenient to consider the following conditions at the wall for the derivatives of mean velocity and Reynolds shear stress [2]

$$\frac{\partial^n \langle uv \rangle}{\partial y^n} = 0, n = 0 \div 2, \frac{\partial^3 \langle uv \rangle}{\partial y^3} = -6\Lambda, \frac{\partial U}{\partial y} = \frac{v_\tau^2}{\nu},$$
$$\frac{\partial^2 U}{\partial y^2} = \frac{\alpha}{\nu}, \frac{\partial^3 U}{\partial y^3} = 0, \frac{\partial^4 U}{\partial y^4} = -\frac{6\Lambda}{\nu}, \qquad (3)$$

where $\Lambda = 0.5 \langle (\partial u/\partial y)(\partial^2 w/\partial y \partial z) \rangle \big|_{y=0}$ is the parameter related to the correlation between the instantaneous streamwise shear and gradient of the transverse shear at the wall. If we choose $l_\nu = v_\tau^{2/3} \Lambda^{-1/3}$ and introduce the new variables $y_\nu = y\Lambda^{1/3} v_\tau^{-2/3}$, $U_\nu = \nu\Lambda^{1/3} U v_\tau^{-8/3}$, $\tau_+ = -\langle uv \rangle / v_\tau^2$, then the equation (1) and expressions (3) take the form

$$\tau_+ = 1 - \frac{\partial U_\nu}{\partial y_\nu} + O(\alpha_\nu y_\nu), \quad \frac{\partial^n \tau_+}{\partial y_\nu^n} = 0, \, n = 0 \div 2, \quad \frac{\partial^3 \tau_+}{\partial y_\nu^3} = 6,$$

$$U_\nu = 0, \frac{\partial U_\nu}{\partial y_\nu} = 1, \frac{\partial^2 U_\nu}{\partial y_\nu^2} = \alpha_\nu, \frac{\partial^3 U}{\partial y_\nu^3} = 0, \frac{\partial^4 U_\nu}{\partial y_\nu^4} = -6 \text{ at } y_\nu = 0. \quad (4)$$

Parameter (4) $\alpha_\nu = \alpha l_\nu v_\tau^{-2}$ is assumed to be small in the limit $\nu \to 0$, since the viscous lenghtscale $l_\nu$ must tend to zero in this limit (pressure gradient assumed to be mild $\alpha\delta v_\tau^{-2} = O(1)$, or smaller). From (4) it follows that in the limit of zero viscosity $\nu \to 0$ the equation and the boundary conditions at the wall for the leading order of the solution in the viscous sublayer have the self-similar form

$$U_\nu = U_\nu(y_\nu), \, \tau_+ = \tau_+(y_\nu). \quad (5)$$

The new scaling (5) gives better collapse for the Reynolds shear stresses than the classical scaling using wall units in the viscous sublayer (Fig.1). Matching inner (5) and outer (2) velocity profiles in the overlap layer $l_\nu \ll y \ll \delta$ gives $U_e + u_o F(\eta) = v_\tau^{8/3} \nu^{-1} \Lambda^{-1/3} f(y_\nu)$, where $F(\eta) = \lim_{\eta \to 0} U_1(\eta; x)$, $f(y_\nu) = \lim_{y_\nu \to \infty} U_\nu(y_\nu)$. Introducing the new variables $\delta_\nu = \delta \Lambda^{1/3} v_\tau^{-2/3}$, $H = \nu \Lambda^{1/3} v_\tau^{-8/3} U_e$, $G = \nu \Lambda^{1/3} v_\tau^{-8/3} u_o$, the latter expression can be rewritten as the functional equation $H(\delta_\nu) + G(\delta_\nu) F(\eta) = f(\eta \delta_\nu)$, which can be easily transformed into Vincze's functional equation [18]. The general solution is

$$f(y_\nu) = A \frac{y_\nu^\gamma - 1}{\gamma} + C, \quad F(\eta) = A \frac{\eta^\gamma - 1}{\gamma} + D, \qquad (6)$$

$$G(\delta_\nu) = \delta_\nu^\gamma, \quad H(\delta_\nu) = A \frac{\delta_\nu^\gamma - 1}{\gamma} + C - D \delta_\nu^\gamma \qquad (7)$$

depending on four arbitrary parameters $A, C, D, \gamma$. Using the traditional wall units ($\nu$ and $v_\tau$) we have $G = u_{o+} \Lambda_+^{1/3}$, $\delta_\nu = \delta_+ \Lambda_+^{1/3}$, where $\Lambda_+ = \Lambda \nu^3 v_\tau^{-5}$, $\delta_+$ is the friction Reynolds number, and the exponent $\gamma$ can be expressed as

$$\gamma = \frac{\ln(u_{o+} \Lambda_+^{1/3})}{\ln(\delta_+ \Lambda_+^{1/3})} = \frac{\ln G}{\ln(\delta_+ \Lambda_+^{1/3})} \qquad (8)$$

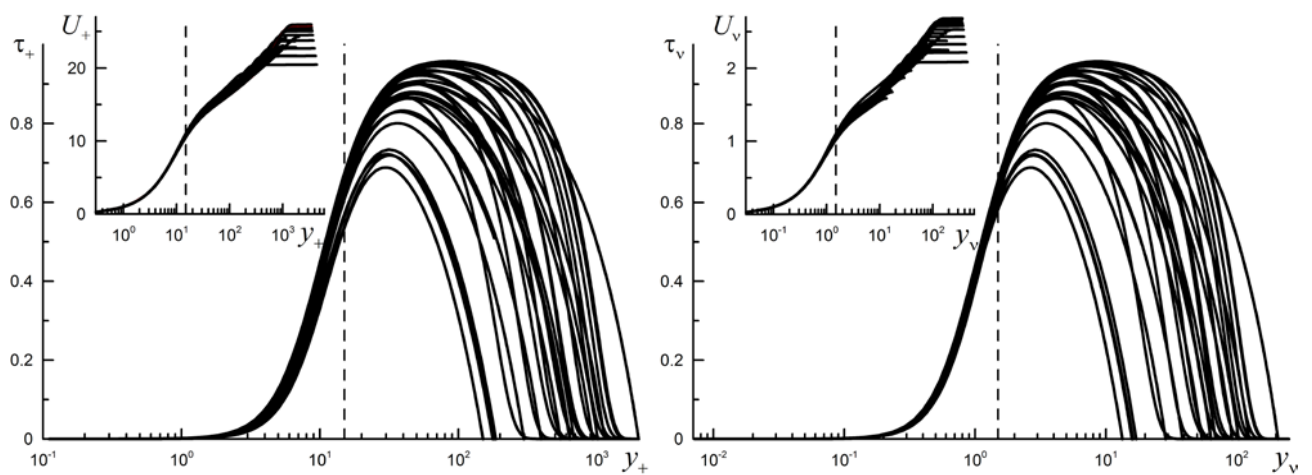

FIG.1. Comparison of classical (left) and new (right) scaling of Reynolds shear stresses and mean velocity (inserts). Curves – DNS data [12–17] in the range $180 < \delta_+ < 2000$, dashed lines mark off the rough border of viscous sublayer.

The behavior of the exponent (8) as $\nu \to 0$, or $\delta_+ \to \infty$, is of particular interest. In the case of complete similarity of wall shear stress statistics one might expect that the criterion $\Lambda_+$ tends to a finite limit when $\delta_+ \to \infty$. The complete similarity can also be assumed for $u_{o+}$ as the outer layer parameter. Then the exponent (8) tends to zero at least as $\ln^{-1} \delta_+$ or faster depending on whether $\ln G$ is $O(1)$ or $G \to 1$. Since $\gamma$ is small, the power-law (6) can be expanded in powers of $\gamma$. In wall units these expansions can be written as

$$U_+ \to \frac{1}{\kappa}\ln y_+ + B + O(\gamma \ln^2 y_+), \quad y_+ \to \infty \qquad (9)$$

$$U_+ - U_{e+} \to \frac{G}{\kappa}\ln \eta + B' + O(\gamma \ln^2 \eta), \quad \eta \to 0 \qquad (10)$$

where $\kappa = \Lambda_+^{1/3} A^{-1}$, $B = \Lambda_+^{-1/3}(C + A \ln \Lambda_+^{1/3})$, $B' = D u_{o+}$. Formally the expression (9) coincides with classical log law, but its coefficients depend on parameter $\Lambda_+$, which can vary with $\delta_+$ (see Fig.2 below). However, the inner limit of the velocity defect (10) coincides with the classical logarithmic asymptote of von Kármán defect law only in the special case $G = 1$. The established validity of the log-law in the limit $\delta_+ \to \infty$ is thus resulted only from the proposition of complete similarity of the criteria $\Lambda_+$ and $u_{o+}$. On the other hand, because the exponent $\gamma$ may approach to zero slowly in the limit $\delta_+ \to \infty$, it may have the non-negligible values for large, but finite, Reynolds numbers. Thus, for finite $\delta_+$ the mean velocity in the overlap region obeys the power-law (6).

Next result of analysis is the non-universality of the power-law exponent $\gamma$. According to (8) it is dependent not only of $\delta_+$, but also of two parameters $u_{o+}, \Lambda_+$. Their values can be influenced by large-scale vortical structures in the outer layer and may be different for channel, pipe and ZPGTBL. Indeed, the most promising candidate for outer velocity scale $u_o$ – Zagarola and Smits scale [7] $u_{ZS}$ – becomes proportional to $v_\tau$ (as in classic scaling) only at $\delta_+ > 10^4$ in Superpipe data and has non-trivial Re-number-dependence at lowest $\delta_+$ [9]. On the other hand, for ZPGTBL the ratio $u_{ZS}/v_\tau$ is nearly constant already for $\delta_+ > 400$ [19]. The difference in behavior of $\Lambda_+$ for different flows is also seen on Fig. 2. Values of $\Lambda_+$ calculated on the base of available DNS data for channel and pipe flows show the certain tendency of increasing with Reynolds numbers, especially for pipe. On the contrary, the values of $\Lambda_+$ for ZPG TBL are nearly constant. We may suppose that for the pipe and channel flow the complete similarity for $u_{o+}$ and $\Lambda_+$ is achieved at far larger Reynolds numbers than for ZPGTBL. It is interesting to note that if we assume the validity of incomplete similarity hypothesis $\Lambda_+ \sim \delta_+^\beta$ and complete similarity for $u_{o+}$, then from (8) follow $\gamma = \beta/(3+\beta)$. Of this type is the correlation $\Lambda_+ = 4 \cdot 10^{-5} \delta_+^{1/2}$ proposed in [20] (dotted curve on Fig.2). Hence the incomplete similarity of $\Lambda_+$ results in the power-law velocity profile (6) with non-vanishing $\gamma$ at $\delta_+ \to \infty$. From Fig.2 it is seen that the pipe data might be the closest candidate for incomplete similarity behavior for presented at Fig.2 range of low and moderate Reynolds numbers. It may explain why the power law was found as a best fit and no log-law was observed in the Superpipe data up to $\delta_+ \sim 10^4$ [7,9].

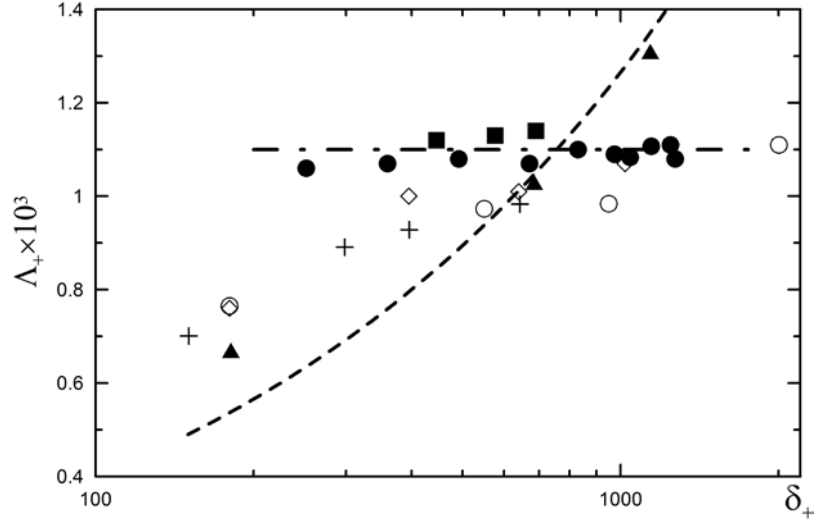

FIG. 2. Parameter $\Lambda_+$ vs. Reynolds number $\delta_+$. DNS of ZPG TBL: [12] (●), [13] (■); pipe [14] (▲); plane channel: [15] (◇), [16](+), [17] (○). Dotted curve – correlation [20].

According to presented theory the power law velocity profile is more general than logarithmic. It is also consistent with the fact, that the power law (7) satisfies the scaling symmetry of inviscid Euler equation of fluid motion

$\mathbf{u}, \mathbf{r}, t \mapsto \lambda^\gamma \mathbf{u}, \lambda \mathbf{r}, \lambda^{1-\gamma} t$ (if additive terms in (6) removed by Galilean transformation) with the particular case $\gamma = 0$ corresponding to the log law [2]. The presented analysis reveals also the absence of universality of $\gamma$ for different types of wall-bounded flows at finite Reynolds numbers owing to observed non-universality of $u_{o+}$ and $\Lambda_+$. The further experimental and DNS studies of the behavior of criteria $u_{o+}$ and $\Lambda_+$ in the wide range of Reynolds numbers and their relationship with the structure of large scales of turbulent motions in wall-bounded flows are merited.

**In summary** our investigation shows that the classical scaling parameters, such as wall units, cannot provide a good collapse of the mean velocity and Reynolds shear stress in the viscous sublayer owing to the variability of the parameter $\Lambda_+$ with the Reynolds number. The new scaling for the viscous sublayer is proposed, which gives the better collapse of available data for both mean velocity and Reynolds shear stresses profiles in the viscous sublayer at a wide range of Reynolds numbers. Derived from the suggested new scaling, the matching condition for mean velocity in the overlap region is equivalent to Vincze's functional equation and has the general power-law solution. Power law exponent tends to zero only in the case of complete similarity of the parameter $\Lambda_+$ in the limit of infinite Reynolds resulting in the classical logarithmic law-of-the-wall. At finite Reynolds numbers the power law exponent is not universal and may be dependent of the flow geometry. The absence of the logarithmic law for the mean velocity observed in previous experiments and direct numerical simulations of turbulent flow in a circular pipe is shown to be resulted from the incomplete similarity of $\Lambda_+$ in a pipe flow. On the other hand the presence of log layer in a turbulent flow in a plane channel and ZPGTBL is connected closely with the observed tendency of the complete similarity of $\Lambda_+$.

The work is supported by the Grant of the RF Government designed to support scientific research projects implemented under the supervision of leading scientists at Russian institutions of higher education (No. 11.G34.31.0046) and RAS Basic Research Programme No.11.